\documentclass[12pt]{article}

\usepackage[a4paper,margin=2.54cm]{geometry}

\usepackage[scaled=.95]{FiraSans}

\usepackage{fancyhdr}
\fancypagestyle{plain}{%
\fancyhf{} % clear all header and footer fields
\fancyfoot[C]{\sffamily\fontsize{11pt}{11pt}\selectfont\thepage} % except the center

}
\usepackage{sectsty}
\allsectionsfont{\sffamily}

\usepackage[most]{tcolorbox}
\usepackage{empheq}

\usepackage{amsmath}
\usepackage{amsfonts} % needed for bold Greek, Fraktur, and blackboard bold
\usepackage{bm}

\usepackage{cite}

\usepackage{hyperref}
\DeclareMathOperator{\Diag}{diag}
\DeclareMathOperator{\Span}{span}
\DeclareMathOperator{\Trace}{tr}
\DeclareMathOperator{\Vect}{vec}

\newcommand{\mcA}{\mathcal{A}}
\newcommand{\mcL}{\mathcal{L}}
\newcommand{\mcT}{\mathcal{T}}

\newcommand{\LG}{\bm{\mathcal{L}}_{+}^\uparrow}
\newcommand{\SL}{SL(2,\mathbb{C})}

\makeatletter
\newcommand*\Cdot{\mathpalette\Cdot@{.5}}
\newcommand*\Cdot@[2]{\mathbin{\vcenter{\hbox{\scalebox{#2}{$\m@th#1\bullet$}}}}}
\makeatother

\begin{document}

\title{\sf\textbf{The Lorentz group and the\\ Kronecker product of matrices}}
% In a long title you can use \\ to force a line break at a certain location.

\author{\sf Jonas Larsson\footnote{\sf\href{mailto:jonas.larsson@umu.se}{jonas.larsson@umu.se}, Dept. Physics, Umeå University, Sweden 
}\qquad \quad Karl Larsson\footnote{\sf\href{mailto:karl.larsson@umu.se}{karl.larsson@umu.se},  Dept. Mathematics and Mathematical Statistics, Umeå University, Sweden}}

\date{\sf\today}

\maketitle % title page is now complete

\begin{abstract}
The group $SL(2,\mathbb{C})$ of all complex $2\times 2$ matrices with determinant one is closely related to the group $\bm{\mathcal{L}}_{+}^\uparrow$ of real $4\times 4$ matrices representing the restricted Lorentz transformations. This relation, sometimes called the spinor map, is of fundamental importance in relativistic quantum mechanics and has applications also in general relativity. In this paper we show how the spinor map may be expressed in terms of pure matrix algebra by including the Kronecker product between matrices in the formalism. The so-obtained formula for the spinor map may be manipulated by matrix algebra and used in the study of Lorentz transformations.
\end{abstract}

\section{Introduction}

In this contribution we use the spinor map for studying Lorentz transformations and the Lorentz group. This map shows that the restricted Lorentz group is closely related to the (apparently much simpler) group of complex $2\times 2$ matrices with determinant one. The spinor map is typically introduced in courses on advanced special relativity, general relativity, relativistic quantum mechanics and quantum field theory \cite{naber,penrose,MR2883916,Maggiore:845116}. It may also be developed from the perspective of Clifford algebra \cite{pertti2001,MR1369094}.  The spinor map is a structure that underlies the existence of spinors which in turn is fundamental for relativistic quantum mechanics. However, while spinors are quite difficult to understand this is not the case with the spinor map and this object can easily be introduced also in undergraduate courses.

An obvious application of the spinor map is to use $2\times 2$ complex
matrices in the study of the real $4\times 4$ matrices that represents the restricted Lorentz transformations. This is attractive because it is much easier to handle $2\times 2$ than $4\times 4$ matrices. By extending standard matrix algebra with the Kronecker product we are able to express the spinor map as a single matrix algebra formula. This turns out to be useful since matrix algebra makes calculations transparent in contrast to alternative formalisms like index algebra, and its simplicity in application makes for efficient and straightforward derivations of basic properties of the Lorentz transformations.

The group of restricted Lorentz transformations\cite{wiki:LorentzGroup,penrose} is represented by a set of real $4\times 4$ matrices that we denote by $\LG$. The closely related group $\SL$ consists of all complex $2\times 2$ matrices with determinant one. A surjective group homomorphism $\SL \to \LG$, sometimes called ``the spinor map''\cite{naber}, may be expressed in terms of the Kronecker matrix product\cite{wiki:KroneckerProduct,steeb}, denoted $\otimes$, as
\begin{empheq}[box={\tcbhighmath[boxsep=0mm, colframe=black, colback=white, boxrule=1.5pt]}]{equation}
\label{eq:spinor-map}
L \to \Lambda_L = \mcA^\dagger \left( L^* \otimes L \right) \mcA
\end{empheq}
Here $\mcA$ is a simple $4\times 4$ matrix (defined in section~\ref{section:spinor-map}), the dagger stands for Hermitian conjugation and the star for complex conjugation. This explicit expression for the spinor map is useful together with matrix algebra.
The 16 elements of the matrix $\Lambda_L$ may be expressed in terms of the 4 elements in
$L = \left[\begin{smallmatrix}a & c \\ b & d \end{smallmatrix}\right]$. Two examples are
\begin{align} \label{eq:Lambda_ab}
\begin{aligned}
\left(\Lambda_L\right)_{00} = \frac{1}{2} \left( a a^* + b b^* + c c^* + d d^* \right)
\\
\left(\Lambda_L\right)_{02} = \frac{i}{2} \left( c a^* - a c^* + d b^* - b d^* \right)
\end{aligned}
\end{align}
and there are 14 more expressions like these. We note that the two components given in equation \eqref{eq:Lambda_ab} are real numbers as they must be. However, for expeditious derivations involving the spinor map the formula \eqref{eq:spinor-map} is more useful  than expressions like \eqref{eq:Lambda_ab}. The reason for this is that \eqref{eq:spinor-map} allows for the use of matrix algebra in a straightforward way. This paper largely consists of derivations of some standard results for Lorentz transformations by use of matrix algebra extended with the Kronecker product of matrices. This makes possible simple derivations accessible also at an undergraduate level.

\paragraph{Related work}
In searching the literature we find only few references where Lorentz transformations in some way are connected with the Kronecker matrix product. There is however one reference with a title and abstract that suggests a close relation to the present paper \cite{sharma}. However, that paper is less elementary than the present one because it deals with linear algebra in a more abstract way. The convenient form \eqref{eq:spinor-map} of the spinor map, that is central to the present paper, is not included in that paper. We expect the present paper to be easily accessible at an undergraduate level while this is not the case with \cite{sharma}. Furthermore, the papers consider different applications where the more mathematical \cite{sharma} considers null cones and projective complex coordinates as well as results for groups like $SU(2)$, $Sp(1)$ and $SO(4)$. The present paper is restricted to matrix formalism and applications of formula \eqref{eq:spinor-map}. In particular, we consider exponential representations of the spinor map, classifications of restricted Lorentz transformations and the generalization of equation \eqref{eq:spinor-map} to non-restricted Lorentz transformations.

\paragraph{Paper outline}
In section~\ref{section:kronecker-product} we consider some basic properties of the Kronecker matrix product. These demonstrate how nicely this product connects with standard matrix algebra making it an easily accessible tool for special relativity. In section~\ref{section:spinor-map} we use this formalism to find a simple derivation of result \eqref{eq:spinor-map}. We prove some basic results for the restricted Lorentz transformations and also consider the exponential form of the spinor map
where $L\in\SL$ is expressed
\begin{align} \label{eq:three}
L = e^\ell
\quad\text{with}\quad
\ell = \begin{bmatrix}
\ell_3 & \ell_1 - i \ell_2 \\ \ell_1 + i \ell_2 & -\ell_3
\end{bmatrix}
\end{align}
Here $\ell_j = \left( -\xi_j + i\omega_j \right)/2$
where $\xi_j,\omega_j$ are six real numbers. Note that $\det(L)=\det\left(e^\ell\right)=e^{\Trace \ell} = e^0 = 1$. The spinor map of $L$ results in the restricted Lorentz transformation (using block matrix notation)
\begin{align} \label{eq:five}
\Lambda_L = e^\lambda
\quad\text{where}\quad
\lambda = - \begin{bmatrix}
0 & \vec\xi^T \\ \vec\xi & \vec\omega \times
\end{bmatrix}
\end{align}
and
\begin{align} \label{eq:six}
\vec\xi = \begin{bmatrix}
\xi_1 \\ \xi_2 \\ \xi_3
\end{bmatrix} \,,
\quad
\vec\omega\times = \begin{bmatrix}
0 & -\omega_3 & \omega_2 \\
\omega_3 & 0 & -\omega_1 \\
-\omega_2 & \omega_1 & 0
\end{bmatrix}
\end{align}
A previous derivation of this simple result is surprisingly complicated and certainly not straightforward, see \cite[section~IIIA]{berk2001}. In contrast, the derivation we obtain by application of formula \eqref{eq:spinor-map} is short and makes use of only simple matrix algebra.

The notation $\LG$ is used for a set of $4\times 4$ real matrices representing the restricted Lorentz transformations. The set of all Lorentz transformations may be written as a union of four disjoint sets\cite{wiki:LorentzTransformation}
\begin{align} \label{eq:seven}
\bm{\mathcal{L}}
=
\bm{\mathcal{L}}^{\uparrow}_+
\cup
\bm{\mathcal{L}}^{\uparrow}_-
\cup
\bm{\mathcal{L}}^{\downarrow}_+
\cup
\bm{\mathcal{L}}^{\downarrow}_-
\end{align}
Here the up or down arrows indicate preservation or reversion of time orientation and the plus or minus signs represent the corresponding for space-time orientation. We in section~\ref{section:kronecker-lorentz-formula} generalize \eqref{eq:spinor-map} also to the non-restricted Lorentz transformations.

In section~\ref{section:classification-restricted-lorentz} we use the Jordan normal form\cite{wiki:JordanNormalForm,basic-algebra} for matrices in $\SL$ together with result \eqref{eq:spinor-map} in order to consider the standard classification of restricted Lorentz transformations. There are four classes of non-identity, restricted Lorentz transformations: \emph{parabolic}, \emph{elliptic}, \emph{hyperbolic} and \emph{loxodromic}\cite{wiki:LorentzGroup}.
More familiar to many readers may be the concepts of a \emph{boost} (i.e. $\vec\omega=0$ in equation \eqref{eq:five}) and a \emph{rotation} (i.e. $\vec\xi=0$ in equation \eqref{eq:five}). A restricted Lorentz transformation may be written as the product of a boost with a rotation. A Lorentz transformation conjugate\cite{wiki:ConjugacyClass,basic-algebra} with a boost is not necessarily a boost and the corresponding is true for rotations. However, a non-identity boost always belongs to the hyperbolic class and a non-identity rotation to the elliptic class.

In section~\ref{section:numerical-example} we consider an explicit example where a real $4\times 4$ matrix $\Lambda$ is given by specifying its 16 elements numerically. Formula \eqref{eq:spinor-map} is used to show with very little calculation that the given $\Lambda$ is in fact a restricted Lorentz transformation. We also find six real parameters $\xi_j$ and $\omega_j$ so that the exponential form \eqref{eq:five} of $\Lambda$ is obtained.

\section{The Kronecker product of matrices}\label{section:kronecker-product}

In this section we define the Kronecker product of matrices and state some basic algebraic rules.
The set of real or complex $m\times n$ matrices is denoted by $\mathbb{R}^{m\times n}$ or $\mathbb{C}^{m\times n}$. In this paper we only consider the Kronecker product of matrices in $\mathbb{C}^{2\times 2}$, but the definitions and algebraic rules presented below many times generalize to matrices of other orders.
We denote the components of $2\times 2$ matrices as
\begin{align}
A = \begin{bmatrix}
A_1 & A_3 \\ A_2 & A_4
\end{bmatrix}
\end{align}
and let $A,B,C,D \in \mathbb{C}^{2\times 2}$ in the remainder of this section.
The Kronecker product of $A$ and $B$ is then a $4\times 4$ matrix defined
\begin{align} \label{eq:nine}
A \otimes B = \begin{bmatrix}
A_1 B & A_3 B \\ A_2 B & A_4 B
\end{bmatrix}
\end{align}
The zero and identity $2\times 2$ matrices are denoted by $0_{2\times 2}$ and $1_{2\times 2}$. We get
\begin{align}
\begin{aligned}
1_{2\times 2} \otimes A
&= \begin{bmatrix}
A & 0_{2\times 2} \\ 0_{2\times 2} & A
\end{bmatrix}
,\ \
A \otimes 1_{2\times 2}
= \begin{bmatrix}
A_1 1_{2\times 2} & A_3 1_{2\times 2} \\ A_2 1_{2\times 2} & A_4 1_{2\times 2}
\end{bmatrix}
,\ \
1_{2\times 2} \otimes 1_{2\times 2} = 1_{4\times 4}
\end{aligned}
\end{align}
Some algebraic rules of the Kronecker product are:
\begin{align}
\label{eq:eleven}
\left( A\otimes B \right)^T &= A^T \otimes B^T
&&\text{(transpose)}
\\
\label{eq:eleven-b}
\left( A\otimes B \right)^{-1} &= A^{-1} \otimes B^{-1}
&&\text{(inverse)}
\\
\label{eq:twelve}
\left( A \otimes B \right) \left( C \otimes D \right)
&= \left( AC \right) \otimes \left( BD \right)
&&\text{(matrix product)}
\\
\label{eq:thirteen}
\det\left( A \otimes B \right) &= \left(\det A\right)^2 \left(\det B\right)^2
&&\text{(determinant)}
\end{align}
Denoting vectorization of a $2\times 2$ matrix as
\begin{align}
\Vect(A) =
\begin{bmatrix}
A_1 & A_2 & A_3 & A_4
\end{bmatrix}^T
\end{align}
we have the relationship
\begin{align} \label{eq:fifteen}
\Vect(ABC) &= \left(C^T\otimes A\right) \Vect(B)
\end{align}
and to the transpose operation we associate a matrix $\mcT$ such that
\begin{align}
\Vect\left(A^T\right) &= \mcT \Vect(A)
\,,\quad
\label{eq:eightteen}
\mcT (A\otimes B) \mcT = B\otimes A
\,,\quad\text{where then}\
\mcT = \left[\begin{smallmatrix}
1 & 0 & 0 & 0 \\
0 & 0 & 1 & 0 \\
0 & 1 & 0 & 0 \\
0 & 0 & 0 & 1
\end{smallmatrix}\right]
\end{align}
In a result below about exponentials we need the Kronecker sum $\oplus$, defined as
\begin{align}
A\oplus B = A \otimes 1_{2\times 2} + 1_{2\times 2} \otimes B
\end{align}
where the matrices $A \otimes 1_{2\times 2}$ and $1_{2\times 2} \otimes B$ commutes (use \eqref{eq:twelve}). It easily follows that
\begin{align} \label{eq:twenty}
e^A \otimes e^B = e^{A\oplus B}
\end{align}

It is straightforward to check most of the relations above by simply writing all expressions in explicit form without using block matrix shorthand notations and just using familiar standard matrix algebra. This procedure may however sometimes be somewhat tedious in spite the fact that we limit our considerations to Kronecker products of $2\times 2$ matrices. For example, the derivation of result \eqref{eq:fifteen} becomes shorter and more transparent if we use block matrix calculations
\begin{align}
\begin{aligned}
\Vect(ABC)
&=
\begin{bmatrix}
ABC \left[\begin{smallmatrix} 1 \\ 0 \end{smallmatrix}\right]
\\
ABC \left[\begin{smallmatrix} 0 \\ 1 \end{smallmatrix}\right]
\end{bmatrix}
%\\&
=
\begin{bmatrix}
C_1 AB \left[\begin{smallmatrix} 1 \\ 0 \end{smallmatrix}\right]
+ C_2 AB \left[\begin{smallmatrix} 0 \\ 1 \end{smallmatrix}\right]
\\
C_3 AB \left[\begin{smallmatrix} 1 \\ 0 \end{smallmatrix}\right]
+ C_4 AB \left[\begin{smallmatrix} 0 \\ 1 \end{smallmatrix}\right]
\end{bmatrix}
\\&=
\begin{bmatrix}
C_1 A & C_2 A \\ C_3 A & C_4 A % error:C4
\end{bmatrix}
\begin{bmatrix}
B \left[\begin{smallmatrix} 1 \\ 0 \end{smallmatrix}\right]
\\
B \left[\begin{smallmatrix} 0 \\ 1 \end{smallmatrix}\right]
\end{bmatrix}
%\\&
=
\left( C^T \otimes A \right) \Vect(B)
\end{aligned}
\end{align}

\section{The spinor map}\label{section:spinor-map}

\subsection{Derivation of the formula for the spinor map}
In this section we derive formula \eqref{eq:spinor-map} for the spinor map in terms of the Kronecker matrix product.
To represent events in space-time we use real 4-vectors ($4\times 1$ matrices) and denote their components
\begin{align}
%x = \begin{bmatrix}x_0 & x_1 & x_2 & x_3 \end{bmatrix}^T \in \mathbb{R}^{4\times 1}
x = \begin{bmatrix}x_0 \\ x_1 \\ x_2 \\ x_3 \end{bmatrix}
= \begin{bmatrix}x_0 \\ \vec x \end{bmatrix}
 \in \mathbb{R}^{4\times 1}
% ,\qquad \vec x = \begin{bmatrix}x_1 \\ x_2 \\ x_3 \end{bmatrix}
\end{align}
The standard reference frame is employed so that for an event $x$ we associate time with $x_0 =ct$ and space with $\vec x $ in Cartesian coordinates, assuming a future direction of $e_0 = \begin{bmatrix}1 & 0 & 0 & 0 \end{bmatrix}^T$ and a right handed orientation of space. For $x,y\in \mathbb{R}^{4\times 1}$ the Lorentz product is defined
\begin{align} \label{eq:twentytwo}
x \Cdot y = x^T \eta y = -x_0 y_0 + x_1 y_1 +x_2 y_2 + x_3 y_3
\end{align}
where the matrix $\eta = \Diag(-1,1,1,1)$.
In terms of the Lorentz product we classify $x$ as time-like if $x\Cdot x < 0$, space-like if $x\Cdot x > 0$ and null if $x\Cdot x = 0$. A Lorentz transformation is represented by a $4\times 4$ matrix $\Lambda$ that satisfies
\begin{align}
\label{eq:lorentz-transformation}
(\Lambda x) \Cdot (\Lambda y) = x\Cdot y \,,\quad \text{i.e. that}\quad
\Lambda^T \eta \Lambda = \eta
\end{align}

\paragraph{Association between vectors and Hermitian matrices}
We associate each 4-vector $x$ with a Hermitian $2\times 2$ matrix $X$ as
\begin{align}
\label{eq:twentyfive}
\begin{aligned}
x \leftrightarrow X &=
\begin{bmatrix}
x_0 + x_3 & x_1 - i x_2 \\ x_1 + i x_2 & x_0 - x_3
\end{bmatrix}
\\&
= x_0 \begin{bmatrix}
1 & 0 \\ 0 & 1
\end{bmatrix}
+ x_1 \begin{bmatrix}
0 & 1 \\ 1 & 0
\end{bmatrix}
+ x_2 \begin{bmatrix}
0 & -i \\ i & 0
\end{bmatrix}
+ x_3 \begin{bmatrix}
1 & 0 \\ 0 & -1
\end{bmatrix}
\\&
= x_0 1_{2\times 2} + x_1 \sigma_1 + x_2 \sigma_2 + x_3 \sigma_3
%\\&
= x_0 1_{2\times 2} + \vec x \cdot \sigma
\end{aligned}
\end{align}
where $\sigma_i$ are the Pauli matrices. Note that
\begin{align}
\det(X) &= -x\Cdot x = x_0^2 - \vec x \cdot \vec x
\end{align}
The relation \eqref{eq:twentyfive} can be expressed
\begin{align}
\label{eq:thirtyone}
\Vect(X) &= \sqrt{2} \mcA x
\quad\text{where}\quad
\mcA = \frac{1}{\sqrt{2}}
\begin{bmatrix}
1 & 0 & 0 & 1 \\
0 & 1 & i & 0 \\
0 & 1 & -i & 0 \\
1 & 0 & 0 & -1
\end{bmatrix}
\end{align}
We include the $\sqrt{2}$ to get the properties
\begin{align} \label{eq:thirtythree}
\mcA^\dagger \mcA = \mcA^T \mcA^* = 1_{4\times 4}
\end{align}

\paragraph{The spinor map}
For $L\in\SL$ we define a map between Hermitian $2\times 2$ matrices $X\to Y$ by
\begin{align} \label{eq:twentyseven}
Y = L X L^\dagger
\end{align}
%For $X$ given by the association to 4-vectors \eqref{eq:twentyfive} this is the so-called spinor map \cite{wiki:LorentzGroup}.
The corresponding map on the associated 4-vectors is
\begin{align}
y = \Lambda_L x
\end{align}
where the real $4\times 4$ matrix $\Lambda_L$ turns out to be a Lorentz transformation, and as we will show in the next section, more specifically a restricted Lorentz transformation. To verify that $\Lambda_L$ is a Lorentz transformation, take the determinant of \eqref{eq:twentyseven} and find $y_0^2 - \vec y \cdot \vec y = x_0^2 - \vec x \cdot \vec x$.
The spinor map $L \to \Lambda_L$ may be written in a more explicit form. By vectorizing both sides of \eqref{eq:twentyseven} and applying equation \eqref{eq:fifteen} we obtain
\begin{align} \label{eq:thirty}
\Vect(Y) = \left( L^* \otimes L \right) \Vect(X)
\end{align}
Expressing $\Vect(X),\Vect(Y)$ in terms of their associated 4-vector $x,y$ via \eqref{eq:thirtyone} and using \eqref{eq:thirtythree} we arrive at
\begin{align} \label{eq:thirtyfour}
y = \mcA^\dagger \left( L^* \otimes L \right) \mcA x
\end{align}
and have thus derived expression \eqref{eq:spinor-map}.

\subsection{The group of restricted Lorentz transformations} \label{section:restricted-lorentz}

In this section we show that the group of restricted Lorentz transformations is constituted by all matrices generated via the spinor map \eqref{eq:spinor-map}, illustrating their intimate relationship.
As apparent by \eqref{eq:lorentz-transformation} the group of all Lorentz transformations may be represented by a set of real $4\times 4$ matrices
\begin{align}
\bm{\mcL} = \left\{ \Lambda \in \mathbb{R}^{4\times 4} \ | \ \Lambda^T \eta \Lambda = \eta \right\}
\end{align}
where $\eta = \Diag(-1,1,1,1)$.
The group of \emph{restricted} Lorentz transformations is defined
\begin{align} \label{eq:thirtyseven}
\bm{\mcL}_+^\uparrow = \left\{ \Lambda \in \bm{\mcL} \ | \ \Lambda_{00}>0\,,\ \det\Lambda >0 \right\}
\end{align}
meaning that a restricted Lorentz transformation in addition to preserving the Lorentz product, also preserves the orientation of time and space-time thanks to the conditions $\Lambda_{00}>0$ and $\det\Lambda >0$. In terms of Lie-group theory this means that $\Lambda$ is continuously connected to the identity $1_{4\times 4}$.

We would like to prove that the group of restricted Lorentz transformations also may be written
\begin{align} \label{eq:thirtyeight}
\bm{\mcL}_+^\uparrow = \left\{ \Lambda_L  = \mcA^\dagger\left(L^* \otimes L \right)\mcA  \ | \ L \in\SL \right\}
\end{align}
We first check that \eqref{eq:thirtyeight} is a matrix group of real matrices. That $\Lambda_L$ is a real matrix follows by use of \eqref{eq:spinor-map} and \eqref{eq:eightteen} as
\begin{align}
\begin{aligned}
\Lambda_L^* &=
\mcA^T \left( L \otimes L^* \right) \mcA^*
\\&=
\mcA^T \mcT \left( L^* \otimes L \right) \mcT \mcA^*
\\&=
\mcA^\dagger \left( L^* \otimes L \right) \mcA
%\\&
= \Lambda_L
\end{aligned}
\end{align}
The group structure also follows because for $L,M \in \SL$ we have by use of \eqref{eq:spinor-map}, \eqref{eq:twelve} and \eqref{eq:thirtythree}
\begin{align}
\Lambda_L \Lambda_M = \Lambda_{LM}
\end{align}
Next, we verify that $\Lambda_L$ is a Lorentz transformation, i.e. we show that
\begin{align} \label{eq:threeeighteen}
\Lambda_L^T \eta \Lambda_L = \eta
\end{align}
From \eqref{eq:spinor-map} we have
\begin{align} \label{eq:fourtytwo}
\Lambda_L^T \eta \Lambda_L = \left[ \mcA^\dagger \left( L^* \otimes L \right) \mcA \right]^T \eta \left[ \mcA^\dagger \left( L^* \otimes L \right) \mcA \right]
\end{align}
and by standard matrix algebra, the algebraic rules \eqref{eq:eleven} and \eqref{eq:twelve}, the identities
\begin{align} \label{eq:fourtythree}
\mcA \eta \mcA^T = \mcA^* \eta \mcA^\dagger = -\varepsilon \otimes \varepsilon
\end{align}
and
\begin{align}
L^T \varepsilon L = L^\dagger \varepsilon L^* = \varepsilon
\quad
\text{where}
\quad
\varepsilon = \begin{bmatrix}
0 & 1 \\ -1 & 0
\end{bmatrix}
\end{align}
we find that \eqref{eq:fourtytwo} can be expressed
\begin{align} \label{eq:three21}
\Lambda_L^T \eta \Lambda_L = -\mcA^T \left[ \varepsilon\otimes\varepsilon \right] \mcA
\end{align}
Applying \eqref{eq:fourtythree} and \eqref{eq:thirtythree} yield the desired result \eqref{eq:threeeighteen}.

The two inequalities  in definition \eqref{eq:thirtyseven} are the conditions for the matrix $\Lambda$ to represent a restricted Lorentz transformation. We now check that $\Lambda_L$ satisfy these and use the notations
\begin{align} \label{eq:fourtyeight}
e_0 = \begin{bmatrix}
1 & 0 & 0 & 0
\end{bmatrix}^T
\quad\text{and}\quad
L = \begin{bmatrix}
a & c \\ b & d
\end{bmatrix}
\end{align}
We have
\begin{align}
\left( \Lambda_L \right)_{00} = e_0^T \Lambda_L e_0 = e_0^T \mcA^\dagger \left( L^* \otimes L \right) \mcA e_0
\end{align}
Since $e_0^T \mcA^\dagger = \left(1/\sqrt{2}\right)\begin{bmatrix}1 & 0 & 0 & 1\end{bmatrix}$ and $\mcA e_0 = \left(1/\sqrt{2}\right)\begin{bmatrix}1 & 0 & 0 & 1\end{bmatrix}^T$ we get
\begin{align}
\left( \Lambda_L \right)_{00} = \frac{1}{2} \left( a a^* + b b^* + c c^* + d d^* \right) > 0
\end{align}
From \eqref{eq:spinor-map}, \eqref{eq:thirteen} and \eqref{eq:thirtythree} we also have
\begin{align} \label{eq:fiftyone}
\det\left[ \Lambda_L \right] = 1 > 0
\end{align}
so $\Lambda_L$ is a restricted Lorentz transformation.

To finish the proof that equations \eqref{eq:thirtyseven} and \eqref{eq:thirtyeight} define the same set of Lorentz transformations we also need the result that for any $\Lambda$ defined by \eqref{eq:thirtyseven} there exists an $L\in\SL$ such that $\Lambda_L = \Lambda$. This follows directly from a general result in Lie group theory. The exact details from Lie group theory is outside the scope of the present paper but in \cite[page 18]{penrose} we find the following explanation. The subgroup $\{\Lambda_L \ | \ L\in \SL \}$ of the restricted Lorentz transformations have full dimensionality six. This is because the matrices $L$ form a six-dimensional (i.e., three-complex-dimensional) system and because only a discrete number (namely two) of the matrices $L$ defines a single restricted Lorentz transformation. It follows that this full-dimensional subgroup must contain the entire connected component of the identity in the Lorentz group.

The result that \eqref{eq:thirtyseven} and \eqref{eq:thirtyeight} define the same set of Lorentz transformations may alternatively be obtained by a direct construction where a restricted Lorentz transformation is written as the product of a boost and a rotation \cite[pages 18--21]{penrose}.

\subsection{The exponential form of the spinor map} \label{section:exponential-form}
In this section we demonstrate how the spinor map \eqref{eq:spinor-map} can be easily rephrased in exponential form.
The equations \eqref{eq:three}--\eqref{eq:six} give the standard exponential form of the spinor map. Both $L\in\SL$ and $\Lambda_L \in \LG$ are expressed in terms of the six real parameters in $\vec\xi$ and $\vec\omega$, associated with boosts and rotations, respectively. By equations \eqref{eq:spinor-map}, \eqref{eq:three}, \eqref{eq:twenty} and \eqref{eq:thirtythree} we easily get $\Lambda_L$ in its exponential form
\begin{align} \label{eq:fiftytwo}
\begin{aligned}
\Lambda_L &= \mcA^\dagger \left( L^* \otimes L \right) \mcA
%\\&
=
\mcA^\dagger \left( e^{\ell^*} \otimes e^{\ell} \right) \mcA
%\\&
=
\mcA^\dagger e^{\ell^* \oplus \ell} \mcA
=
e^{\mcA^\dagger\left( \ell^* \oplus \ell \right) \mcA}
\end{aligned}
\end{align}
It remains to be shown that the exponent is the correct one, i.e.
\begin{align}
\mcA^\dagger\left( \ell^* \oplus \ell \right) \mcA = \lambda
\end{align}
We have
\begin{align}
\begin{aligned}
\ell^* \oplus \ell
&=
\ell^* \otimes 1_{2\times 2} + 1_{2\times 2}\otimes\ell
\\&=
\begin{bmatrix}
\ell_3^* & \ell_1^* + i \ell_2^* \\ \ell_1^* - i \ell_2^* & -\ell_3^*
\end{bmatrix} \otimes 1_{2\times 2}
%\\&\quad
+ 1_{2\times 2}\otimes\begin{bmatrix}
\ell_3 & \ell_1 - i \ell_2 \\ \ell_1 + i \ell_2 & -\ell_3
\end{bmatrix}
\\&=
\begin{bmatrix}
\ell_3 + \ell_3^* & \ell_1 - i \ell_2 & \ell_1^* + i \ell_2^* & 0 \\
\ell_1 + i \ell_2 & \ell_3^* - \ell_3 & 0 & \ell_1^* + i \ell_2^* \\
\ell_1^* - i \ell_2^* & 0 & \ell_3 - \ell_3^* & \ell_1 - i \ell_2 \\
0 & \ell_1^* - i \ell_2^* & \ell_1 + i \ell_2 & -\ell_3 - \ell_3^*
\end{bmatrix}
\end{aligned}
\end{align}
which gives us that
\begin{align}
&\mcA^\dagger\left( \ell^* \oplus \ell \right) \mcA
=
%\\ &\ \nonumber
\begin{bmatrix}
0 & \left (\ell_1 + \ell_1^* \right) & \left (\ell_2 + \ell_2^* \right) & \left (\ell_3 + \ell_3^* \right)
 \\
\left (\ell_1 + \ell_1^* \right) & 0 & -i\left(\ell_3 - \ell_3^* \right) & i \left (\ell_2 - \ell_2^* \right) \\
\left (\ell_2 + \ell_2^* \right) & i\left(\ell_3 - \ell_3^* \right) & 0 & -i \left(\ell_1 - \ell_1^* \right) \\
\left(\ell_3 + \ell_3^* \right) & -i \left (\ell_2 - \ell_2^* \right) & i \left(\ell_1 - \ell_1^* \right) & 0
\end{bmatrix}
\end{align}
Substituting $\ell_j = \left(-\xi_j + i\omega_j\right)/2$ we find
\begin{align} \label{eq:fiftysix}
\begin{aligned}
\mcA^\dagger\left( \ell^* \oplus \ell \right) \mcA
&=
\begin{bmatrix}
0 & -\xi_1 & -\xi_2 & -\xi_3 \\
-\xi_1 & 0 & \omega_3 & -\omega_2 \\
-\xi_2 & -\omega_3 & 0 & \omega_1 \\
-\xi_3 & \omega_2 & -\omega_1 & 0
\end{bmatrix}
%\\&
=
- \begin{bmatrix}
0 & \vec\xi^T \\ \vec\xi & \vec\omega\times
\end{bmatrix}
\end{aligned}
\end{align}
This is the exponent $\lambda$ in equation \eqref{eq:five}.

\section{Formulas for general Lorentz transformations}\label{section:kronecker-lorentz-formula}

In this section we generalize formula \eqref{eq:spinor-map} for the spinor map to general Lorentz transformations.
A Lorentz transformation belongs to one of four components as expressed by equation \eqref{eq:seven}. For a restricted Lorentz transformation we already have the formula \eqref{eq:spinor-map} but we now include an arrow and a plus-sign in the notation
\begin{align} \label{eq:fiftyseven}
\Lambda_L^{\uparrow+} = \mcA^\dagger \left( L^* \otimes L \right) \mcA
\end{align}
where the arrow pointing upwards represents preservation of the direction of time and the plus sign represents the preservation of space-time orientation. A procedure to derive the matrix \eqref{eq:fiftyseven} is given by equations \eqref{eq:twentyseven}--\eqref{eq:thirtyfour}. A direct proof that the matrix so defined is an element in $\LG$ is also given in section~\ref{section:spinor-map}. If we start from
\begin{align}
Y = LX^T L^\dagger
\end{align}
rather than from \eqref{eq:twentyseven} we get in place of \eqref{eq:fiftyseven}
\begin{align} \label{eq:fiftynine}
\Lambda_L^{\uparrow-} = \mcA^\dagger \left( L^* \otimes L \right) \mcA^*
\end{align}
where we used $\mcT \mcA = \mcA^*$.
We may easily check that $\Lambda_L^{\uparrow-}\in\bm{\mcL}_-^\uparrow$ by procedures similar to those we used in section~\ref{section:spinor-map} to prove that $\Lambda_L^{\uparrow+}\in\LG$.
The elements in $\bm{\mcL}_+^\downarrow$ and $\bm{\mcL}_-^\downarrow$ may be written
\begin{align} \label{eq:sixtyone}
\Lambda_L^{\downarrow+} = -\Lambda_L^{\uparrow+}
\,,\quad
\Lambda_L^{\downarrow-} = -\Lambda_L^{\uparrow-}
\end{align}
The product of any two elements in $\bm{\mcL}$ may be obtained from
\begin{align}
\begin{alignedat}{2}
\Lambda_L^{\uparrow+} \Lambda_M^{\uparrow+} &= \Lambda_{LM}^{\uparrow+} \,,\quad&
\Lambda_L^{\uparrow+} \Lambda_M^{\uparrow-} &= \Lambda_{LM}^{\uparrow-} \,,
\\
\Lambda_L^{\uparrow-} \Lambda_M^{\uparrow+} &= \Lambda_{LM^*}^{\uparrow-} \,,\quad&
\Lambda_L^{\uparrow-} \Lambda_M^{\uparrow-} &= \Lambda_{LM^*}^{\uparrow+}
\end{alignedat}
\end{align}
together with \eqref{eq:sixtyone}.

\paragraph{Simple examples}
As a first example using the formulas above, we let $L=1_{2\times 2}$, which gives the Lorentz transformations
\begin{align}
\begin{alignedat}{2}
\Lambda_L^{\uparrow+}  &= 1_{4\times 4} \,,\quad&
\Lambda_L^{\uparrow-}  &= \Diag(1,1,-1,1) \,,
\\
\Lambda_L^{\downarrow+} &= -1_{4\times 4} \,,\quad&
\Lambda_L^{\downarrow-} &= \Diag(-1,-1,1,-1)
\end{alignedat}
\end{align}
As a second example, we let $L=\varepsilon$, which gives the Lorentz transformations
\begin{align}
\begin{alignedat}{2}
\Lambda_L^{\uparrow+}  &= \Diag(1,-1,1,-1) \,,\quad&
\Lambda_L^{\uparrow-}  &= -\eta \,,
\\
\Lambda_L^{\downarrow+} &= \Diag(-1,1,-1,1) \,,\quad&
\Lambda_L^{\downarrow-} &= \eta
\end{alignedat}
\end{align}

\section{Classification of restricted Lorentz transformations}\label{section:classification-restricted-lorentz}

\subsection{Parabolic, elliptic, hyperbolic and loxodromic transformations}
A non-identity restricted Lorentz transformation $\Lambda_L$ may be classified into the four types \emph{parabolic}, \emph{elliptic}, \emph{hyperbolic} and \emph{loxodromic} \cite{wiki:LorentzGroup}.
The type is determined by the value of $(\Trace L)^2$ where $(\Trace L)^2=4$ in the parabolic case, $0 \leq (\Trace L)^2 < 4$ in the elliptic case, $(\Trace L)^2 > 4$ in the hyperbolic case, and $(\Trace L)^2 \in \mathbb{C}\setminus[0,4]$ in the loxodromic case. The hyperbolic class is thus a subclass of the loxodromic and the loxodromic class consists of all transformations $\Lambda_L$ that are not parabolic or elliptic.
This classification is obtained from the theory of Möbius transformations because the restricted Lorentz group is isomorphic to the Möbius group $PSL(2,\mathbb{C})$. We refer the reader to the references \cite{wiki:MobiusTransformation, olsen2010, gareth-singerman} for the relations between Lorentz and Möbius transformations (a discussion of these relations is outside the scope of the present paper).

Let us now use the spinor map in the form \eqref{eq:spinor-map} as well as the Jordan normal form for matrices in $\SL$. The characteristic polynomial of $L$ may be written
\begin{align}
\begin{aligned}
p_L(s) &= \det\left( L - s 1_{2\times 2} \right)
\\&=
s^2 - \left(\Trace L\right) s + 1
=
(s-s_1)(s-s_2)
\end{aligned}
\end{align}
where $\Trace L = s_1 + s_2$ and $\det L = s_1 s_2 = 1$. There are two cases:

\paragraph{Case I $\bm{\mathsf{(s_1\neq s_2)}}$.} This is the non-parabolic case, i.e. $\left(\Trace L\right)^2\neq 4$.
By application of the Jordan normal form we have
\begin{align}
L \sim L_{I} =
\begin{bmatrix}
s_1 & 0 \\ 0 & s_2
\end{bmatrix}
\end{align}
where the equivalence relation is matrix similarity \cite{wiki:MatrixSimilarity,basic-algebra}
(in this case the same as $\SL$ group conjugacy \cite{wiki:ConjugacyClass,basic-algebra}).
We also have the $\LG$ group conjugacy relation
\begin{align} \label{eq:group-conjugacy}
\Lambda_L \sim \Lambda_{L_I}
\end{align}
Hence, there exists $M\in\SL$ such that
\begin{align} \label{eq:seventyfour}
\Lambda_L = \Lambda_M \Lambda_{L_I} \Lambda_{M^{-1}}
\end{align}
We can express
\begin{align}
s_1 = e^{(\beta + i\theta)/2}
\quad\text{and}\quad
s_2 = e^{-(\beta + i\theta)/2}
\end{align}
with real $\beta$ and $\theta$ such that
\begin{align}
e^{(\beta + i\theta)/2} \neq 1
\end{align}
and thus
\begin{align} \label{eq:seventyseven}
L_I^* \otimes L_I =
\begin{bmatrix}
e^{\beta} & 0 & 0 & 0 \\
0 & e^{-i\theta} & 0 & 0 \\
0 & 0 & e^{i\theta} & 0 \\
0 & 0 & 0 & e^{-\beta}
\end{bmatrix}
\end{align}
From \eqref{eq:spinor-map} together with \eqref{eq:thirtythree}, \eqref{eq:seventyseven} and \eqref{eq:group-conjugacy} it follows that the matrices $\Lambda_L$, $\Lambda_{L_I}$ and $L_I^* \otimes L_I$ are similar. Thus, they have the same characteristic polynomial
\begin{align} \label{eq:charpol}
\begin{aligned}
p_{\Lambda_L}(s) &= p_{\Lambda_{L_I}}(s)
%\\&
=
\left( s - e^{\beta} \right)
\left( s - e^{-\beta} \right)
\left( s - e^{i\theta} \right)
\left( s - e^{-i\theta} \right)
\end{aligned}
\end{align}
We also get from \eqref{eq:spinor-map} and \eqref{eq:seventyseven} that
\begin{align} \label{eq:fivenine}
\Lambda_{L_I} &=
\begin{bmatrix}
\cosh\beta & 0 & 0 & \sinh \beta \\
0 & \cos\theta & \sin\theta & 0 \\
0 & -\sin\theta & \cos\theta & 0 \\
\sinh\beta & 0 & 0 & \cosh \beta
\end{bmatrix}
\end{align}
The traces of $L$ and $L_I$ are the same so
\begin{align}
\begin{aligned}
\left(\Trace L\right)^2 &= (s_1 + s_2)^2
%\\&
=
2 + 2\cosh\beta \cos\theta + 2i\sinh\beta\sin\theta
\end{aligned}
\end{align}
In the elliptic case $\beta=0$ and $\exp(i\theta)\neq 1$ so that $\left(\Trace L\right)^2 = 2 + 2\cos\theta < 4$. In the hyperbolic case $\exp(i\theta)=1$ and $\beta\neq 0$ so that $\left(\Trace L\right)^2 = 2 + 2\cosh\beta > 4$.

\paragraph{Case II $\bm{\mathsf{(s_1= s_2)}}$.}
This is the parabolic case $\left(\Trace L\right)^2=4$ so that $\Trace L=\pm 2$. Since $\Lambda_L = \Lambda_{-L}$ it is sufficient to consider the $\Trace L=2$ case. Then by application of the Jordan normal form
\begin{align} \label{eq:fiveeleven}
\Lambda_L \sim \Lambda_{L_{II}}
\quad\text{for}\quad
L_{II} = \begin{bmatrix}
1 & 1 \\ 0 & 1
\end{bmatrix}
\end{align}
We get in the same way as in \emph{Case I}
\begin{align}
p_{\Lambda_{L_{II}}}(s) = (s-1)^4
\end{align}
and
\begin{align} \label{eq:fivethirteen}
\Lambda_{L_{II}} = \begin{bmatrix}
3/2 & 1 & 0 & -1/2 \\
1 & 1 & 0 & -1 \\
0 & 0 & 1 & 0 \\
1/2 & 1 & 0 & 1/2
\end{bmatrix}
\end{align}

This classification of a restricted Lorentz transformation may be defined directly in terms of its characteristic polynomial \eqref{eq:charpol}.
Then $\Lambda$ is elliptic if $\beta=0$ and $\exp(i\theta)\neq 1$, hyperbolic  if $\beta\neq 0$ and $\exp(i\theta)= 1$, loxodromic  if $\beta\neq 0$ and parabolic if $\beta=0$, $\exp(i\theta)= 1$ and $\Lambda \neq 1_{4\times 4}$.

Some understanding of the meanings of the classification of $\Lambda$ is obtained by considering the following space of 4-vectors
\begin{align}
\mathbb{I}_{\Lambda} = \left\{ x\in\mathbb{R}^{4\times 1} \ |\ \Lambda x =x\right\}
\end{align}
along with its orthogonal complement
\begin{align}
\mathbb{I}_{\Lambda}^\perp = \left\{ x\in\mathbb{R}^{4\times 1} \ |\ x\Cdot y =0 \ \text{for all $y\in\mathbb{I}_{\Lambda}$} \right\}
\end{align}
From \eqref{eq:group-conjugacy} and \eqref{eq:fivenine} we find that if $\Lambda$ is elliptic or hyperbolic then $\mathbb{I}_{\Lambda}$ is a space with dimension 2 and
\begin{align}
\mathbb{I}_{\Lambda}
\cap
\mathbb{I}_{\Lambda}^\perp
=
\left\{
\begin{bmatrix}
0 & 0 & 0 &0
\end{bmatrix}^T
\right\}
\end{align}
so that
\begin{align}
\mathbb{R}^{4\times 1} = \mathbb{I}_{\Lambda} \oplus \mathbb{I}_{\Lambda}^\perp
\end{align}
where the usual direct sum $\oplus$ appears (not the Kronecker sum). An elliptic or hyperbolic $\Lambda$ is ``non-trivial'' only on the 2D space $\mathbb{I}_{\Lambda}^\perp$. In the elliptic case $\mathbb{I}_{\Lambda}^\perp$ contains only space-like 4-vectors while in the hyperbolic case it contains also time-like 4-vectors. In the loxodromic but non-hyperbolic case the dimension of $\mathbb{I}_{\Lambda}$ is zero so $\Lambda$ is non-trivial on the full 4D space. In the parabolic case we use \eqref{eq:fiveeleven} and \eqref{eq:fivethirteen} and find that the dimension of $\mathbb{I}_{\Lambda}$ is two and
\begin{align}
\mathbb{I}_{\Lambda} \cap \mathbb{I}_{\Lambda}^\perp = \Span\left\{ n \right\}
\end{align}
where $n$ is a non-zero null vector.

\subsection{Boost and rotations}

We classify a non-identity Lorentz transformation $\Lambda_L$ as a \emph{boost} if $\vec\omega =\vec 0$, i.e.
\begin{align}
L = \exp\left[-\frac{1}{2}\vec\xi\cdot\sigma\right] \neq 1_{2\times 2}
\end{align}
which can be expressed
\begin{align}
L = \cosh\left(\xi/2\right) 1_{2\times 2} - \sinh\left(\xi/2\right) \hat\xi\cdot\sigma
\end{align}
where $\xi = \sqrt{\vec\xi\cdot\vec\xi} > 0$ and $\hat\xi = \vec\xi/\xi$. We get
\begin{align}
\left(\Trace L\right)^2 = 4\left[ \cosh\left(\xi/2\right) \right]^2 > 4
\end{align}
so a boost is hyperbolic. Note that $L^\dagger = L$, so $L$ is Hermitian for a boost.

We classify a non-identity Lorentz transformation $\Lambda_L$ as a \emph{rotation} if $\vec\xi =\vec 0$, i.e.
\begin{align}
L = \exp\left[\frac{i}{2}\vec\omega\cdot\sigma\right] \neq 1_{2\times 2}
\end{align}
which can be expressed
\begin{align}
L = \cos\left(\omega/2\right) 1_{2\times 2} + i\sin\left(\omega/2\right) \hat\omega\cdot\sigma
\end{align}
where $\omega = \sqrt{\vec\omega\cdot\vec\omega} >0$ and $\hat\omega = \vec\omega/\omega$.
We get
\begin{align}
\left(\Trace L\right)^2 = 4\left[ \cos\left(\omega/2\right) \right]^2 < 4
\end{align}
so a rotation is elliptic. Note that $L^\dagger = L^{-1}$, so $L$ is unitary for a rotation.

\section{A numerical example}\label{section:numerical-example}

Consider the matrix
\begin{align} \label{eq:ninety}
\Lambda =
\begin{bmatrix}
9/4 & 0 & 1/4 & 2 \\
0 & 1 & 0 & 0 \\
1 & 0 & 1 & 1 \\
7/4 & 0 & -1/4 & 2
\end{bmatrix}
\end{align}
We will now; (a) prove that $\Lambda$ is a restricted Lorentz transformation, (b) determine what class $\Lambda$ belongs to, and (c) write $\Lambda$ in exponential form.

\paragraph{The matrix is a restricted Lorentz transform}
We prove that $\Lambda$ is a restricted Lorentz transformation by writing it in the form of $\Lambda_L$ in equation \eqref{eq:spinor-map}. Rewriting \eqref{eq:spinor-map} as
\begin{align} \label{eq:ninetyone}
L^* \otimes L = \mcA \Lambda \mcA^\dagger
\end{align}
we by straightforward algebra get
\begin{align} \label{eq:ninetytwo}
L^* \otimes L &=
\begin{bmatrix}
4 & 0 & 0 & 0 \\
i & 1 & 0 & 0 \\
-i & 0 & 1 & 0 \\
1/4 & -i/4 & i/4 & 1/4
\end{bmatrix}
\end{align}
By inspection of \eqref{eq:ninetytwo}, definition \eqref{eq:nine} and using $\det L = 1$ we easily find
\begin{align} \label{eq:ninetythree}
L = \pm\begin{bmatrix}
2 & 0 \\ i/2 & 1/2
\end{bmatrix}
\end{align}
We have now in a very economical way proved that $\Lambda$ is a restricted Lorentz transformation. We also note that $\left(\Trace L\right)^2 = 25/4 > 4$, so $\Lambda$ is hyperbolic.

\paragraph{The matrix in exponential form}
To write $\Lambda$ in exponential form, we first note that we may write \eqref{eq:three} as
\begin{align}
L = e^\ell = e^{\vec\ell \cdot \sigma}
\end{align}
where we use notations from \eqref{eq:twentyfive}, $\vec\ell = \bigl(-\vec\xi + i\vec\omega\bigr)/2$ and
\begin{align} \label{eq:ninetyfive}
\ell^2 = ( \vec\ell \cdot \vec\ell ) 1_{2\times 2}
\end{align}
From the power series defining the exponential of a matrix and equation \eqref{eq:ninetyfive} we find
\begin{align} \label{eq:ninetysix}
L = \cosh(\mu) 1_{2\times 2} + \frac{\sinh(\mu)}{\mu} \vec\ell \cdot \sigma
\end{align}
Here we use the notation $\mu=\sqrt{\vec\ell\cdot\vec\ell}=\bigl\{\alpha\in\mathbb{C}\ | \ \alpha^2 = \vec\ell\cdot\vec\ell \bigr\}$. This is a set with two elements such that $\alpha\in\mu \Rightarrow -\alpha\in\mu$. We define $\cosh(\mu)=\cosh(\alpha)=\cosh(-\alpha)$ and $\mu^{-1}\sinh(\mu)=\alpha^{-1}\sinh(\alpha)=(-\alpha)^{-1}\sinh(-\alpha)$. The two branches of the square root are included and there is no need to choose branch. We write $L$ in the form
\begin{align}
L = \frac{\Trace L}{2} 1_{2\times 2} + \vec L \cdot \sigma
\end{align}
where (choosing the plus sign in \eqref{eq:ninetythree})
\begin{align} \label{eq:ninetyeight}
\Trace L = \frac{5}{2} \,,\quad \vec L = \frac{1}{4}\begin{bmatrix}
i & 1 & 3
\end{bmatrix}^T
\end{align}
From \eqref{eq:ninetysix}--\eqref{eq:ninetyeight} we get
\begin{align}
\cosh(\mu) = \frac{5}{4} \,,\quad \frac{\sinh(\mu)}{\mu}\vec\ell = \frac{1}{4}\begin{bmatrix}
i & 1 & 3
\end{bmatrix}^T
\end{align}
A solution to the first equation is $\mu=\{\pm\ln 2\}$ and we get
\begin{align}
\vec\ell = \frac{\ln 2}{3}
\begin{bmatrix}
i & 1 & 3
\end{bmatrix}^T
\end{align}
and thus
\begin{align}
\vec\xi = -\frac{2\ln 2}{3}\begin{bmatrix}
0 & 1 & 3
\end{bmatrix}^T
\,,\quad
\vec\omega = \frac{2\ln 2}{3}\begin{bmatrix}
1 & 0 & 0
\end{bmatrix}^T
\end{align}
An exponential expression for $\Lambda$ is now given by \eqref{eq:five}.\\

\section{Discussion}\label{section:discussion}

The spinor map is a powerful tool for the construction and investigation of Lorentz transformations. It takes $2 \times 2$ complex matrices representing the group $\SL$ to $4\times 4$ real matrices representing the Lorentz transformations $\LG$. Since the spinor map is a 2-to-1 group homomorphism it tell us a lot about Lorentz transformations in terms of the group $\SL$. For simplicity one would like to use matrix algebra and, as far as possible, avoid explicit use of the matrix components in the applications of the spinor map. This is in particular the case at the undergraduate level where the students are familiar with matrix algebra but not much with index formalism.
However, relations involving both $2\times 2$ and $4\times 4$ matrices are not straightforward to describe using standard matrix algebra.
A solution to this problem is to extend the matrix formalism by including the Kronecker product between $2\times 2$ matrices. This is not at all difficult and, as seen in section~\ref{section:kronecker-product}, the Kronecker product very nicely fits into standard matrix algebra. We obtain the convenient formula \eqref{eq:spinor-map} for the spinor map involving $2\times 2$ and $4\times 4$ matrices as well as the Kronecker product. 

This paper consists of applications of formula \eqref{eq:spinor-map} using the matrix algebra extended with the Kronecker product. One example when a derivation is much simplified by our procedure concerns the exponential form of the spinor map. This is obvious by comparing section~\ref{section:exponential-form} with the corresponding derivation in \cite[section IIIA]{berk2001}. We also show how to modify formula \eqref{eq:spinor-map} in order to obtain also the non-restricted Lorentz transformations (section~\ref{section:kronecker-lorentz-formula}). Formula \eqref{eq:spinor-map} is convenient for considering the conjugacy classes (elliptic, hyperbolic, loxodromic and parabolic) of the restricted Lorentz transformations (section~\ref{section:classification-restricted-lorentz}).  %We also note in section~\ref{section:numerical-example} how a certain ``hidden'' symmetry of the spinor map may be used to show, with very little algebra, that a given, real $4\times 4$ matrix represents a restricted Lorentz transformation. 

The matrix algebra approach in the present paper makes the theory of the spinor map more accessible also at an undergraduate level. The extension of standard matrix algebra by the inclusion of the Kronecker product of matrices is worthwhile both because of its simplicity and of applications not only in relativity but also in quantum mechanics \cite{fernandez2016}.

%%%%%%%%%%%%%%%%%%%%%%%%%%%%%%%%%%%%%%%%%%%%

%\bibliographystyle{habbrv}
\bibliographystyle{hunsrt}
{
\footnotesize\sf
\bibliography{lorentz-group-refs}
}

\end{document}